\documentclass{article}

\usepackage[a4paper, margin=1in]{geometry}
\usepackage[utf8]{inputenc}
\usepackage{amsmath}
\usepackage{graphicx}
\usepackage{natbib}
\usepackage{bbm}

\usepackage{setspace}
\title{Estimating survival parameters under conditionally independent left truncation}
\author{Arjun Sondhi \\ Flatiron Health}
\date{\today}

\begin{document}

\maketitle

\begin{abstract}
Databases derived from electronic health records (EHRs) are commonly subject to left truncation, a type of selection bias induced due to patients needing to survive long enough to satisfy certain entry criteria. 
Standard methods to adjust for left truncation bias rely on an assumption of marginal independence between entry and survival times, which may not always be satisfied in practice. 
In this work, we examine how a weaker assumption of conditional independence can result in unbiased estimation of common statistical parameters. 
In particular, we show the estimability of conditional parameters in a truncated dataset, and of marginal parameters that leverage reference data containing non-truncated data on confounders. 
The latter is complementary to observational causal inference methodology applied to real world external comparators, which is a common use case for real world databases. 
We implement our proposed methods in simulation studies, demonstrating unbiased estimation and valid statistical inference. 
We also illustrate estimation of a survival distribution under conditionally independent left truncation in a real world clinico-genomic database.
\end{abstract}

\section{Introduction}

In time-to-event analyses, the outcome variable of interest is defined as the time from an initiating event to a terminal event, which is often subject to censoring. 
Time-to-event data may additionally be truncated, meaning that subjects whose time-to-event falls outside a certain interval cannot be observed \citep{klein2003survival}. 
Truncation is commonly found in healthcare data collected outside of clinical trials, such as databases derived from electronic health records (EHRs), where patients need to satisfy certain entry criteria in order to be observed \citep{agarwala2018real}. 
Then, when analyzing survival, any patients who died before satisfying the entry criteria (e.g. undergoing a biomarker testing procedure) are not observed; this is known as left truncation, and is a selection bias since patients observed in the database had to live long enough to qualify for entry \citep{chubak2013threats, cain2011bias}.

Given left truncated data, standard methods for estimating the marginal survival distribution are instead estimating survival in the target population conditional on surviving up to entry time.
Applying risk set adjustment, where patients are only counted at risk for death once they have satisfied the entry criteria, is the most common approach to estimate the marginal distribution.
This can be implemented with both Kaplan-Meier survival estimators and Cox proportional hazards regression models \citep{tsai1987note}. 
However, risk set adjustment relies on the assumption of independence between the time to event $T$ and the time to entry $E$ (both measured from a common initiating event). 
This assumption is known as independent left truncation, and is testable from observed truncated data \citep{mackenzie2012survival, martin2005testing}. 

In practice, the left truncation mechanism is not always independent. 
Consider an EHR-derived database of patients with cancer who undergo comprehensive genomic profiling (CGP). 
This is known as real world data (RWD), collected observationally outside of a clinical trial. 
If patients are tested later in their treatment course due to worsening of disease and exhaustion of standard therapies, this would result in dependency between the survival and entry time (i.e. time to CGP). 
Given this dependent left truncation, risk set adjustment is no longer guaranteed to unbiasedly recover the survival distribution of interest. 
Although some proposed methods claim to estimate marginal survival in the presence of dependent left truncation, they make strong parametric assumptions that are not testable from observed data.
They also generally do not extend to multivariate regression modeling \citep{chaieb2006estimating}.

In this work, we consider the setting of conditionally independent left truncation, where survival time $T$ and entry time $E$ are independent conditional on a set of confounding variables $Z$. 
This is weaker than the marginally independent left truncation assumption described above, and can also be easily tested. 
Intuitively, this is similar to missing at random (MAR) data, where the probability of an observation being missing is independent of its true value, conditional on a set of observed variables. 
Likewise, data where an observation’s true value directly affects its missingness probability (missing not at random or non-ignorable missingness) is analogous to dependently truncated data \citep{bhaskaran2014difference}. 
We show that under conditional independence of $T$ and $E$ given $Z$, unbiased estimates of certain survival parameters can be obtained. 
Our methods are motivated by common analyses performed with real world data from EHR-derived sources. 

We first consider estimation of survival regression parameters, by observing that the conditional survival distribution $T \mid Z$ in the target population is estimable. 
Then, by adjusting for confounders $Z$, a Cox model’s regression parameters can be estimated and interpreted as the conditional log hazard ratios in the correct target population. 
In order to extend this to estimating parameters of the marginal survival distribution $T$, we require additional reference data on the marginal confounder distribution $Z$, e.g. non-truncated observations from the target population. 
Note that the existence of a non-truncated dataset would not remove the need to conduct estimation in the truncated dataset if it is of interest to analyze non-truncated data (such as from a clinical trial) and truncated data together, e.g. to compare their health outcomes. 
We consider the setting of comparing a external cohort from RWD to a single-arm treated cohort from a clinical trial. 
The latter group may be treated with a novel therapy, and the former with standard of care treatments. 
By weighing the external cohort to match the confounder distribution in the trial cohort, we show that the causal hazard ratio comparing the treatment to the control can be estimated. 
This weighting procedure corresponds to that used in estimation of the average treatment effect on the treated (ATT). 
Using a similar method, we can also estimate the marginal survival distribution in a dependently truncated cohort by weighting towards a reference dataset with confounder variables observed. 
This requires the additional assumption that the reference distribution corresponds to the target population of interest, which is true by definition for the ATT.  

The rest of this paper is organized as follows. 
We first detail our methodology for estimating survival parameters under dependent left truncation.
We then demonstrate in simulation studies that our methods can consistently estimate the parameters of interest. 
Finally, we illustrate estimation of a marginal survival distribution in an analysis of data from a real world EHR-derived database, and conclude with a short discussion.

\section{Methods for conditionally independent left truncation}

\subsection{Background}

As defined previously, let $T$ denote the survival time, $E$ denote the entry time, and $C$ denote the censoring time from a common initiating event. 
Then, let $Y = \min(T, C)$ denote the observed time-to-event, and $\delta = I(T \le C)$ indicate whether a death event was observed. 
We assume throughout that $T$ and $C$ are independent. 
The observed data then consists of $(Y_i, \delta_i, E_i) \mid Y_i > E_i, i = 1, \dots, n$. 
With these variables observed, it is only possible to determine whether $T$ and $E$ are marginally independent or dependent. 
Under independent left truncation, applying risk set adjustment to a Kaplan-Meier analysis can recover the true marginal distribution of $T$. 

Now suppose a set of confounding variables $Z$ are also observed at baseline, such that $T$ is independent of $E$ conditional on $Z$. 
Throughout this paper, we define `baseline' to be the time of the initiating event or start of follow-up from which the times $T$ and $E$ are measured.
For example, if $Z$ is a single categorical variable (e.g. sex), then among patients with a particular $Z = z$, survival and entry times are independent. 
As such, applying risk set adjustment in a single strata $z$ would consistently estimate the distribution $T \mid Z = z$, without the truncation selection bias $Y > E$. 
More generally, under conditionally independent left truncation, the distribution $T \mid Z$ is estimable from data.
We exploit this property in order to estimate various survival parameters in the true target population of interest. 

As with marginally independent left truncation, it is possible to test for conditionally independent left truncation. 
This can be done by fitting a risk set adjusted Cox proportional hazards model, with entry time $E$ and confounders $Z$ as covariates. 
If the coefficient for entry time is significantly different from the null value, this provides evidence that the survival and entry times are dependent, conditional on confounders. 

\subsection{Estimating conditional hazard ratios}

\begin{figure}
    \centering
    \includegraphics[scale=0.45]{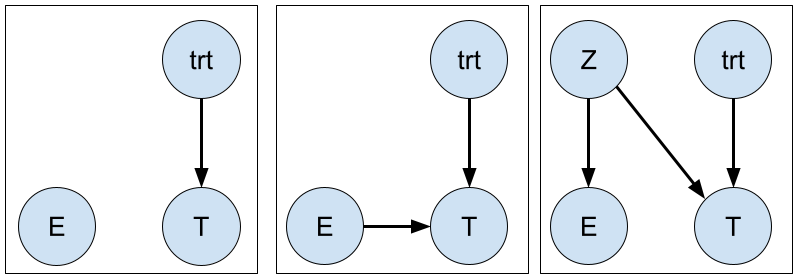}
    \caption{Graphical illustrations of independent (left), dependent (centre), and conditionally independent (right) left truncation.}
    \label{fig:graphs}
\end{figure}

Suppose we are interested in estimating the covariate-adjusted effect of a binary treatment, denoted by the variable $trt$, on survival. 
Causal graphs corresponding to different possible left truncation scenarios are shown in Figure~\ref{fig:graphs}.

Given conditionally independent left truncation, we can fit the Cox proportional hazards model:
$$
\lambda_T(t | trt, Z) = \lambda_{bh}(t) \exp(\beta \; trt + \beta_Z ' Z),
$$
where $\lambda_{bh}(t)$ is an unspecified baseline hazard function.
Then, the parameter $\beta$ corresponds to:
$$
\exp(\beta) = \cfrac{\lambda_T(t \mid trt = 1, Z)}{\lambda_T(t \mid trt = 0, Z)},
$$
which is a ratio of two functionals of the conditional distribution $T \mid Z$.
Because this distribution can be estimated from our observed data, we can also estimate the log hazard ratio $\beta$ unbiasedly. 
For simplicity in Figure~\ref{fig:graphs}, $trt$ and $Z$ are shown as independent.
However, estimability would still hold given an edge from $Z$ to $trt$, because $E$ and $T$ would remain conditionally independent in that scenario. 
This also extends to regression parameters of any other covariates $X$ that may be included in the model, as long as conditioning on them would not induce an association between entry and survival times. 
Practically, we expect most variables of interest that are adjusted for would be collected at baseline (at the time of the initiating event, or start of follow-up), and therefore are unlikely to result in this association. 

\subsection{Estimating marginal hazard ratios and survival distributions}

We have demonstrated how conditionally independent left truncation can yield unbiased estimation of certain conditional hazard ratio parameters using the conditional distribution $T \mid Z$. 
In practice, however, marginal parameters that describe population-averaged effects are often of primary interest. 
In order to estimate these, we require the marginal survival distribution $T$. 
We can express the density of $T$ as:
$$
\pi_T(t) = \int_z \pi_{T \mid Z}(t|z) \; \pi_Z(z) \; dz,
$$
which provides insight into how to recover the marginal distribution. 
The first component $T | Z$ is estimable from observed data; however, in our truncated dataset, we only observe $Z \mid Y > E$, rather than the marginal distribution $Z$. 
Now suppose we are able to obtain weights $w = \frac{\pi(z)}{\pi(z | y > e)}$ corresponding to the density ratio between $Z$ and $Z \mid Y > E$. 
We can then weigh our observed data distribution to match the true marginal distribution:
\begin{align*}
    \pi(y, e, z \mid y > e) \; w 
    &= \pi(y, e \mid z, y > e)\pi(z \mid y > e) \; w \\
    &= \pi(y, e \mid z, y > e)\pi(z) \\
    &= \pi(y \mid z, y > e)\pi(e \mid z, y > e)\pi(z) \\
    &= \pi(y \mid z)\pi(e \mid z)\pi(z) \\
    &= \pi(y, e \mid z)\pi(z) \\
    &= \pi(y, e, z)
\end{align*}
In the above decomposition, the third and fourth equalities follow as a consequence of conditionally independent left truncation. 
Intuitively, we want to weigh the distribution of $Z$ in our sample to make it similar to the correct target population distribution. 

In order to estimate these weights, we require a reference sample of $Z$ drawn from a population that is non-truncated, and therefore representative. 
Then, the density ratio can be estimated through the following procedure:
\begin{enumerate}
    \item Vertically concatenate the non-truncated ($Z$) and truncated ($Z | Y > E$) confounder observations.
    \item Label each observation $J = 1$ if it is from the non-truncated dataset or $J = 0$ if it is from the truncated dataset.
    \item Train a probabilistic classifier that estimates $P(J = 1 \mid Z)$.
    \item For a patient in the truncated dataset having confounders $Z = z$, the estimated weight is then $\hat{w}_i = \cfrac{\hat{P}(J = 1 \mid Z = z)}{\hat{P}(J = 0 \mid Z = z)} \cfrac{\hat{P}(J = 0)}{\hat{P}(J = 1)}$, which targets the density ratio $\cfrac{\pi(z)}{\pi(z | y > e)}$.
\end{enumerate}
This result follows from a simple application of Bayes rule \citep{sugiyama2012density, sondhi2020balanced}. 
In practice, software such as the \texttt{WeightIt} package in R \citep{greifer2019package} can compute the term $\frac{\hat{P}(J = 1 \mid Z = z)}{\hat{P}(J = 0 \mid Z = z)}$ by estimating balancing weights targeting the average treatment effect on the treated (ATT).
Here, patients in the non-truncated sample have weights set to 1.

Although access to a reference sample may seem like a strong condition for estimation of marginal parameters, this naturally occurs in analyses where truncated data is analyzed together with a non-truncated data source. 
An example of this is comparing survival of patients in a single-arm clinical trial to that of real-world patients as an external comparator arm \citep{carrigan2019evaluation, davies2018comparative, carrigan2020using}. 
This may be done to demonstrate treatment effectiveness, or to contextualize early-phase results for rare diseases or biomarkers when enrolling a trial control arm can be difficult. 
Here, the trial arm is not subject to left truncation, while the real world arm potentially is. 
Given conditionally independent left truncation, we can weigh the real world data towards the confounder distribution in the trial arm as described above. 
Then, fitting a weighted and risk set adjusted Cox proportional hazards model with the arm indicator as the sole covariate will estimate a marginal hazard ratio comparing the trial treatment and real world cohorts. 
The corresponding log partial likelihood is:
\begin{align*}
    \ell(\beta) = \sum_{i : \delta_i = 1} \hat{w}_i \Bigg[ \beta \; trt_i - \log \Bigg( \sum_{j \in R_i} \exp(\beta \; trt_j) \Bigg) \Bigg],
\end{align*}
where $\hat{w}_i$ is the estimated weight for patient $i$, and $R_i$ denotes the set of patients in the risk set at the observed event time $T_i$. 
In fact, this is the same procedure used to estimate the ATT hazard ratio (which requires the causal assumption that $Z$ contains sufficient information to satisfy treatment assignment ignorability). 
In other words, we have shown that standard causal inference methodology can provide unbiased estimation when applied to left truncated data, given conditional independence of survival and entry times.

For other analyses, it may be of interest to estimate the marginal survival distribution in a real world dataset subject to left truncation, without making any statistical comparison. 
Under conditionally independent left truncation, this can be done in a similar manner by estimating weights given a reference dataset as above. 
This requires an additional assumption that the reference confounder distribution is representative of the target population of interest, which is definitionally true when estimating the ATT. 
The weights can then be used by fitting a weighted and risk set adjusted Kaplan-Meier estimator for the survival distribution.
Here, the conditional probability of failure at time $x_j$ is estimated as:
\begin{align*}
    \hat{F}(x_{j})
    &= \cfrac{\sum_{i=1}^n I(E_i \le x_j, Y_i = x_j) \; \delta_i \; \hat{w}_i}{ \sum_{i=1}^n I(E_i \le x_j \le Y_i) \; \hat{w}_i},
\end{align*}
where $\hat{w}_i$ is the estimated weight for subject $i$.
A consistency proof of this estimator is sketched in the appendix. 

Statistical inference for the ATT estimated by the weighted Cox proportional hazards model should be done using robust standard errors (which are fit by default when the \texttt{coxph} function in R is given weights), since a weighted analysis induces dependence among observations. 
For the same reason, inference for the marginal survival distribution estimated by the weighted Kaplan-Meier estimator can be done by bootstrapping from the weighted empirical distribution.

\section{Simulation studies}

\subsection{Design}

We simulate a non-truncated treatment arm (such as from a single-arm clinical trial) and a real world control arm that is subject to conditionally independent left truncation. 
First, we generate two confounding variables as:
\begin{align*}
    Z_1 &\sim Bernoulli(1 - p_E) \\
    Z_2 &\sim Normal(0, 0.5),
\end{align*}
where $p_E$ is the probability of entering the cohort at or before the start of follow-up (i.e. without delayed entry) in the real world arm. We set this parameter to be 0.2. Then, cohort entry times for the real world patients are generated as:
\begin{align*}
    E &\sim 
    \begin{cases} 
      0 & Z_1 = 0 \\
      Exponential(\lambda_{entry}) & Z_1 = 1 \\
   \end{cases} \\
    \lambda_{entry} &= \lambda_{ebh} \exp(\beta_{entry} Z_2),
\end{align*}
indicating that patients with $Z_1 = 0$ do not have delayed entry (i.e. an entry time of 0). 
The constant parameter $\lambda_{ebh}$ is set via root-finding in order to achieve a certain truncation probability, given all the other parameters.
We vary the truncation probability $P(Y > E | trt = 0)$ from 0.1 to 0.7 by increments of 0.1.
The parameter $\beta_{entry}$ is varied among $\log(0.5, 0.8, 0.1)$; for $\beta_{entry} \neq 0$, patients with higher values of $Z_2$ are more likely to enter the cohort later.
The cohort entry times for the non-truncated patients are all 0.

We use the following model for survival times among two randomly assigned treatment groups:
\begin{align*}
    trt &\sim Bernoulli(0.5) \\
    T &\sim Exponential(\lambda_{T}) \\
    \lambda_{T} &= \lambda_{bh} \exp(\beta \; trt + \beta_Z \; Z_1 + \beta_Z \; Z_2),
\end{align*}
where the parameters were set as follows:
\begin{itemize}
    \item We set $\lambda_{bh}$ to be 1/12, which corresponds to an average survival time of 12 months for patients on the $trt = 0$ arm with confounders $Z_1 = Z_2 = 0$.
    \item The parameter $\beta$ was set to $\log(0.8)$, which implies that patients on the $trt = 1$ arm have 80\% of the hazard of death as do the patients on the $trt = 0$ arm with the same confounder values. 
    \item $\beta_Z$ is the parameter describing the association between survival and both confounders. We varied this parameter among $\log(1, 1.5, 2)$; for $\beta_Z \neq 0$, patients with higher values of $Z_1$ and $Z_2$ are more likely to die sooner.
\end{itemize}
This design considers a positive association between the confounders $Z$ and the hazard of death combined with a negative association between $Z$ and the hazard of entry.
This results in a spurious marginal association of decreased survival with later time to entry, which is commonly observed in our motivating application of real world clinico-genomic data; patients who are genomic tested later in their disease course tend to have worse survival.

Finally, the censoring model used was 
\begin{align*}
    C &\sim Exponential(\lambda_{C}) \\
    \lambda_{C} &= \lambda_{T},
\end{align*}
corresponding to a censoring probability of 0.5 in the complete dataset (before truncation is applied).
Given this data-generating process, the observed time-to-event is $Y = \min(T, C)$.
Truncation is applied by filtering out observations where $Y < E$ in the real world $trt = 0$ arm. 

We fix the expected real world arm sample size at 250 by generating a larger dataset with sample size $\frac{250}{P(Y > E | trt = 0)}$ and applying the truncation. 
The non-truncated treatment arm has a fixed sample size of 250.
For each parameter configuration, we conduct 1,000 simulation iterations.

\subsection{Conditional parameters}

\begin{figure}
    \centering
    \includegraphics[scale=0.25]{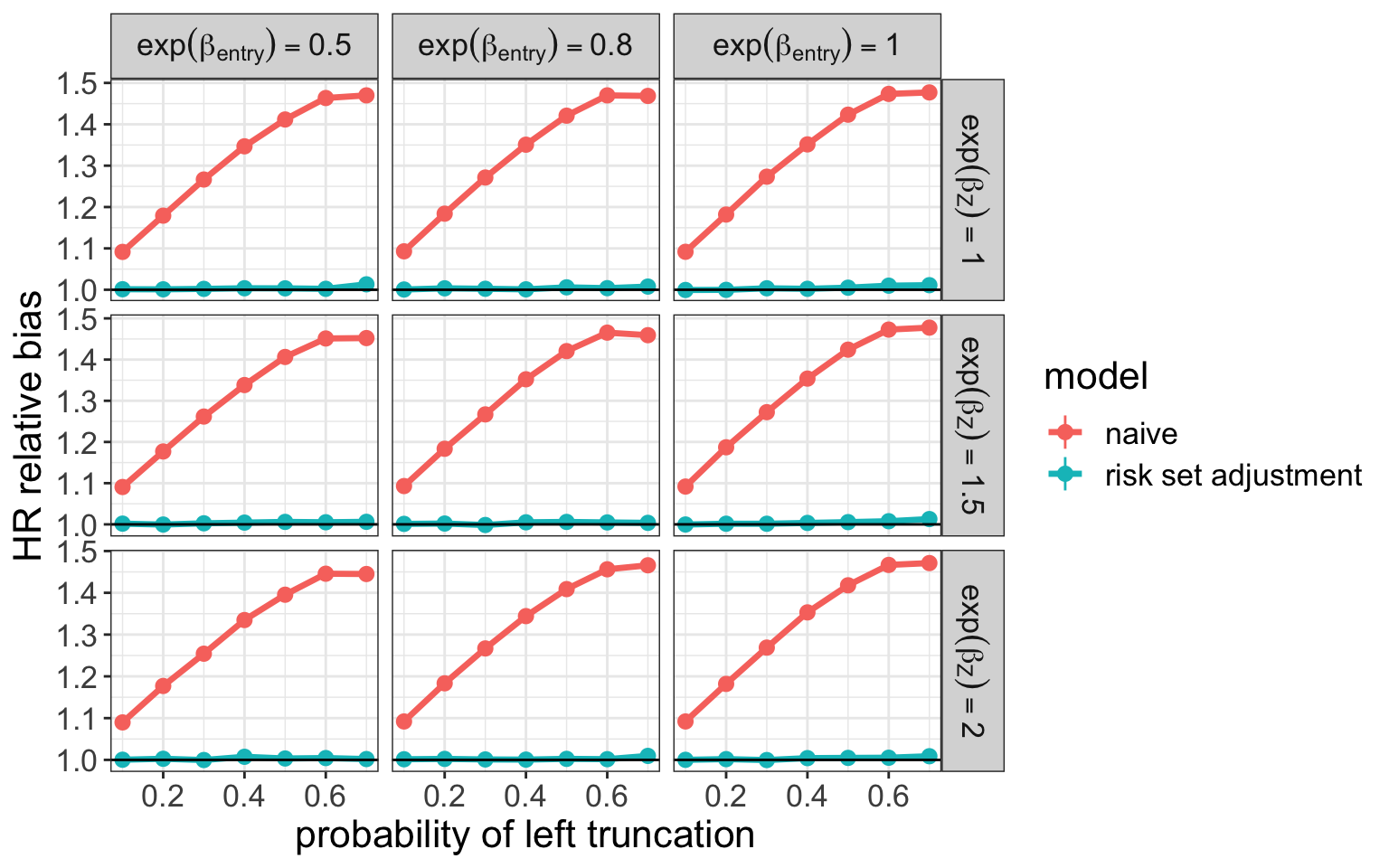}
    \caption{Relative bias for estimated conditional hazard ratio comparing non-truncated arm to real world treatment arm across simulation settings.}
    \label{fig:cond_bias}
\end{figure}

\begin{figure}
    \centering
    \includegraphics[scale=0.25]{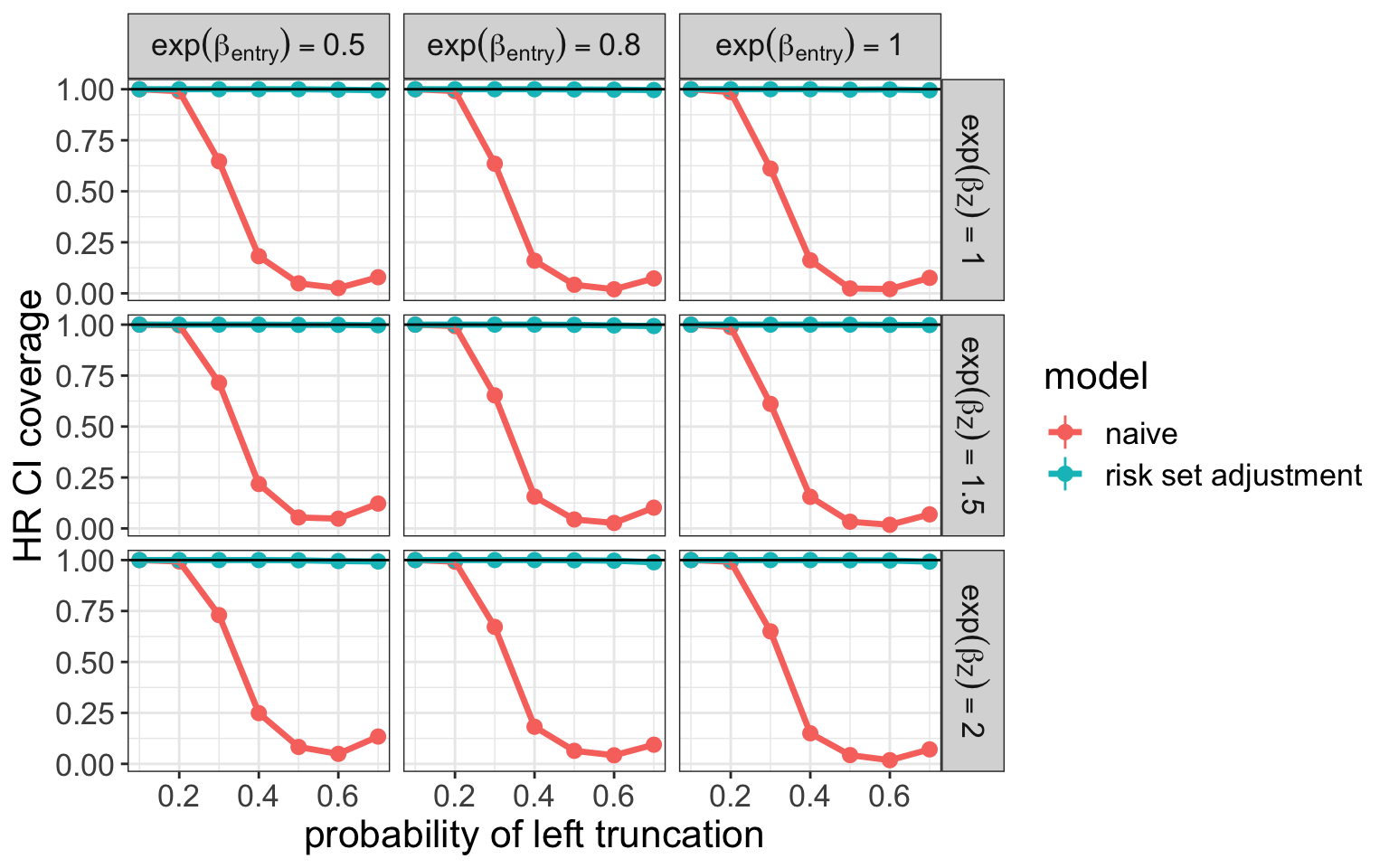}
    \caption{95\% confidence interval coverage for estimated conditional hazard ratio comparing non-truncated arm to real world treatment arm across simulation settings.}
    \label{fig:cond_cov}
\end{figure}

Using the simulation framework described above, we demonstrate how the conditional hazard ratio comparing treatment arms can be estimated unbiasedly under conditionally independent left truncation. 
In each simulation iteration, we estimated this parameter on the complete dataset without truncation applied by fitting a Cox proportional hazards model adjusting for the treatment arm and the confounders; this estimate was considered the ground truth. 
We then fit the same model on the dataset where the real world arm was left truncated, with and without risk set adjustment. 
To evaluate the performance of these estimators, we report the relative bias of the hazard ratio estimate with respect to the true hazard ratio (Figure~\ref{fig:cond_bias}), and coverage of the 95\% confidence intervals (Figure~\ref{fig:cond_cov}). 
We see that the risk set adjusted Cox estimator is able to provide unbiased estimates and valid coverage, despite the marginal dependence between entry time and survival that is induced through the confounding variables. 
As expected, the naive model fit without adjusting for delayed entry results in severe bias. 
In general, this is unaffected by the relationship between survival and the other variables, as shown by similar results for all parameter configurations. 

\subsection{Marginal parameters}

\begin{figure}
    \centering
    \includegraphics[scale=0.25]{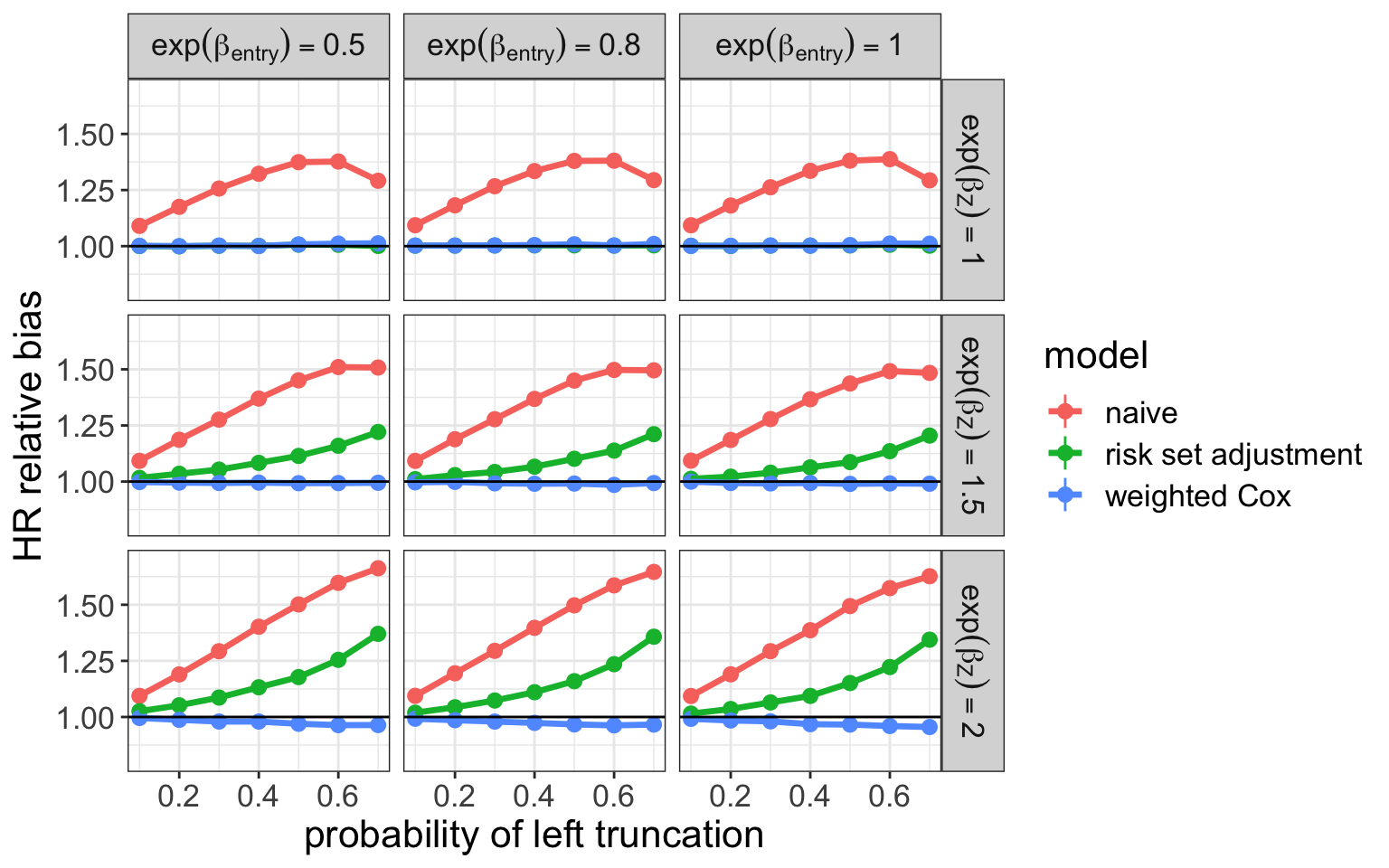}
    \caption{Relative bias for estimated marginal hazard ratio comparing non-truncated arm to real world treatment arm across simulation settings.}
    \label{fig:marg_hr_bias}
\end{figure}

\begin{figure}
    \centering
    \includegraphics[scale=0.25]{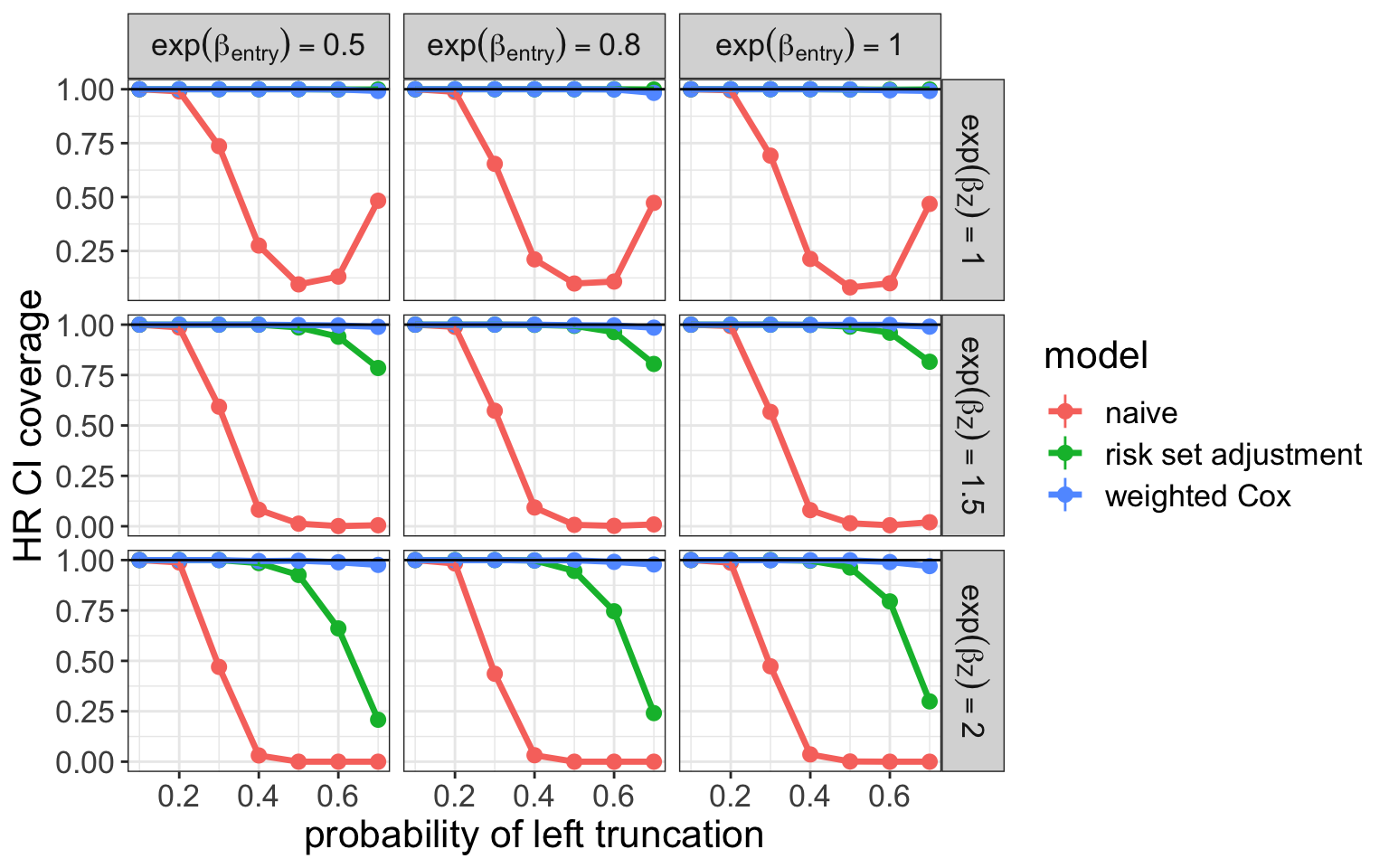}
    \caption{95\% confidence interval coverage for estimated marginal hazard ratio comparing non-truncated arm to real world treatment arm across simulation settings.}
    \label{fig:marg_hr_cov}
\end{figure}

We now demonstrate our proposed methodology for estimating the marginal hazard ratio comparing the non-truncated and truncated arms. 
In each simulation iteration, we estimated this parameter on the complete dataset without truncation applied. 
This estimate was treated as the ground truth, and compared to those produced under the following methods applied to the truncated dataset: 
\begin{itemize}
    \item \textbf{naive:} Cox model estimator that does not account for delayed entry 
    \item \textbf{risk set adjustment:} Cox model estimator that accounts for delayed entry 
    \item \textbf{weighted Cox:} Risk set adjusted Cox model estimator with real world arm weighted towards confounder distribution in non-truncated arm
\end{itemize}
To evaluate the performance of these estimators, we examine the relative bias of the hazard ratio estimate with respect to the true hazard ratio. 
We also examine the coverage of the associated 95\% confidence intervals. 
The results are shown in Figures~\ref{fig:marg_hr_bias} and \ref{fig:marg_hr_cov}. 
We see that the weighted Cox model estimator has low bias and maintains valid coverage at all levels of truncation. 
When $\beta_Z = 0$, both the risk set adjusted and weighted Cox models are unbiased, since $Z_1$ and $Z_2$ do not affect survival, and are therefore not confounders. 
As the strength of confounding increases, the risk set adjusted Cox estimator shows increasing bias and loses valid coverage under high truncation probability.

\begin{figure}
    \centering
    \includegraphics[scale=0.25]{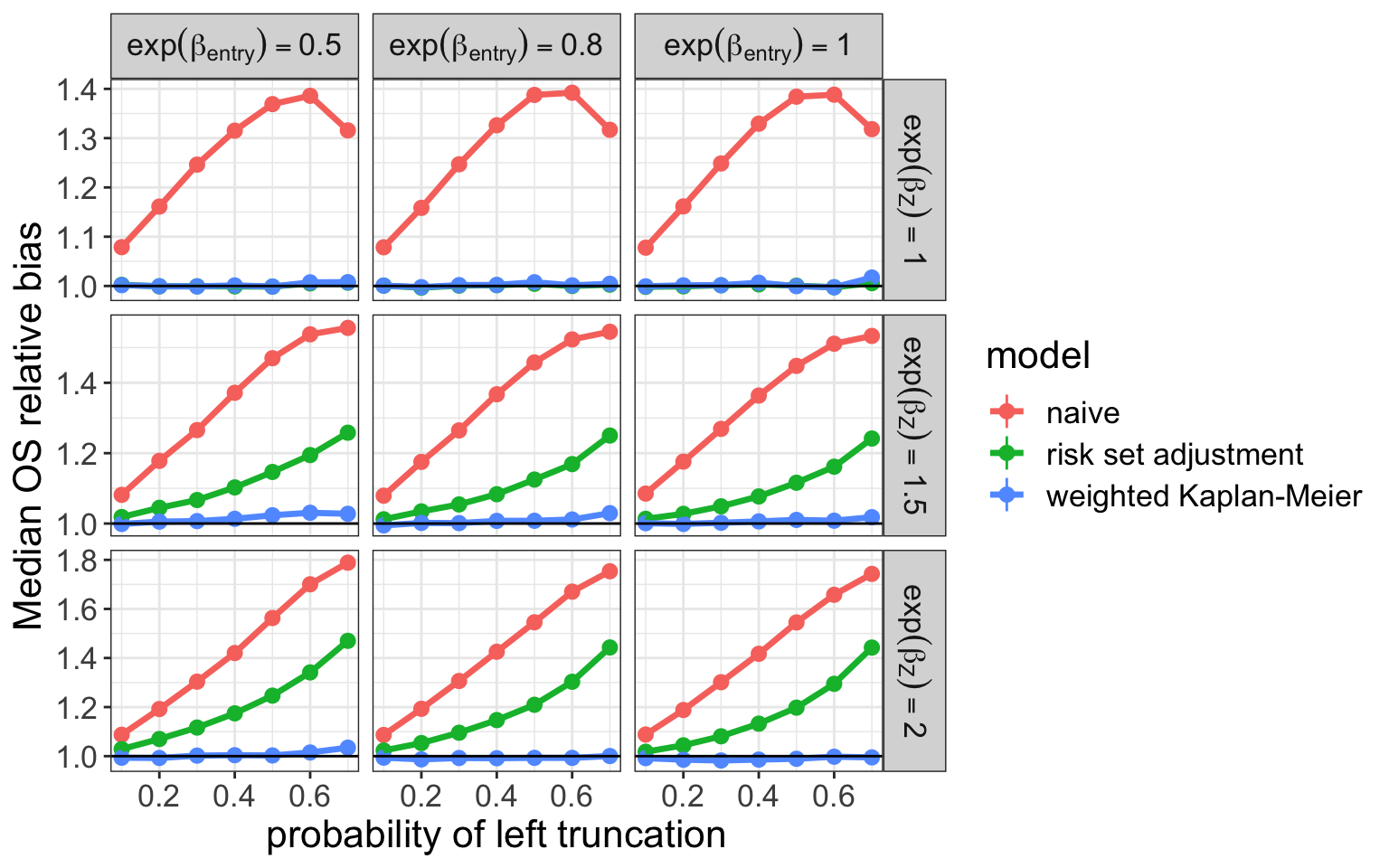}
    \caption{Relative bias for estimated median survival time in real world treatment arm across simulation settings.}
    \label{fig:marg_bias}
\end{figure}

We also examine estimation of the marginal survival distribution in the real world arm. 
In each simulation iteration, we estimated the median survival time on the complete real world cohort. 
This estimate was treated as the ground truth, and compared to those produced under the following methods applied to the truncated cohort: 
\begin{itemize}
    \item \textbf{naive:} Kaplan-Meier estimator that does not account for delayed entry 
    \item \textbf{risk set adjustment:} Kaplan-Meier estimator that accounts for delayed entry 
    \item \textbf{weighted Cox:} Risk set adjusted Kaplan-Meier estimator with real world arm weighted towards confounder distribution in non-truncated arm
\end{itemize}
To evaluate the performance of these estimators, we examine the relative bias of the median survival time estimate, shown in Figure~\ref{fig:marg_bias}. 
As with the hazard ratio, we see that the weighted estimator has low bias at all levels of truncation, while the unweighted risk set adjusted estimator is biased in the presence of confounding.

\section{Application: overall survival in NGS-tested patients with metastatic prostate cancer}

We apply the weighting method described here to estimate the survival distribution of patients with metastatic prostate cancer who received a Foundation Medicine next generation sequencing (NGS) test, taken from the nationwide (US-based) de-identified Flatiron Health-Foundation Medicine Clinico-Genomic Database (CGDB). 
The de-identified data originated from approximately 280 US cancer clinics (representing around 800 sites of care). 
Retrospective longitudinal clinical data were derived from electronic health record (EHR) data, comprising patient-level structured and unstructured data, curated via technology-enabled abstraction, and were linked to genomic data derived from FMI comprehensive genomic profiling (CGP) tests in the FH-FMI CGDB by de-identified, deterministic matching \citep{singal2019association}. 
Genomic alterations were identified via comprehensive CGP of over 300 cancer-related genes on Foundation Medicine's next-generation sequencing (NGS) test \citep{frampton2013development}.

We are specifically interested in estimating overall survival time from the start of first-line (1L) therapy in the metastatic setting. 
The NGS test report date is considered the cohort entry date, so a patient who was tested after their 1L start date would have delayed entry. 
Patients who died before being tested are left truncated from the cohort. 
The observed event time for each patient is either their date of death or date of last follow-up activity (at which the patient is censored). 
Patients with a gap of greater than 90 days between their metastatic diagnosis date and their first subsequent structured EHR activity are excluded from the analysis. 
This resulted in a final cohort size of 1,256.

For our initial analysis, we fit a risk set adjusted Kaplan-Meier estimator to the observed entry, death, and censoring times. 
This yields a median overall survival time of 17.8 months post-1L therapy start, with a 95\% confidence interval of [16.5, 19.3]. 
We then test for marginally dependent left truncation. 
This is done by fitting a risk set adjusted Cox proportional hazards model to estimate the association between survival time and time to NGS testing from 1L therapy start. 
We estimate a hazard ratio of 1.01 with 95\% confidence interval [1.0, 1.02] (p-value = 0.0075).
This indicates a small degree of dependent left truncation; patients are estimated to have a 1\% greater hazard of death for each additional later month from 1L start to NGS testing. 
Given this dependence, the median survival estimate given above may be inaccurate, though the bias would likely be low, since the magnitude of the association is small.

Next, we look to see if this dependence can be explained by observed baseline covariates; these were selected a priori from among the variables measured in both databases. 
We fit a risk set adjusted Cox model of overall survival on entry time that controls for year of metastatic diagnosis, clinic practice type (academic vs community), patient’s age at metastatic diagnosis, patient’s race, tumour histology, and group stage at diagnosis.
We estimate a conditional hazard ratio of 1.00 with 95\% confidence interval [0.99, 1.01] (p-value = 0.17), providing evidence that left truncation is independent conditional on these covariates.
Therefore, weighting to the appropriate distribution of these variables would recover the correct survival distribution. 

\begin{figure}
    \centering
    \includegraphics[scale=0.2]{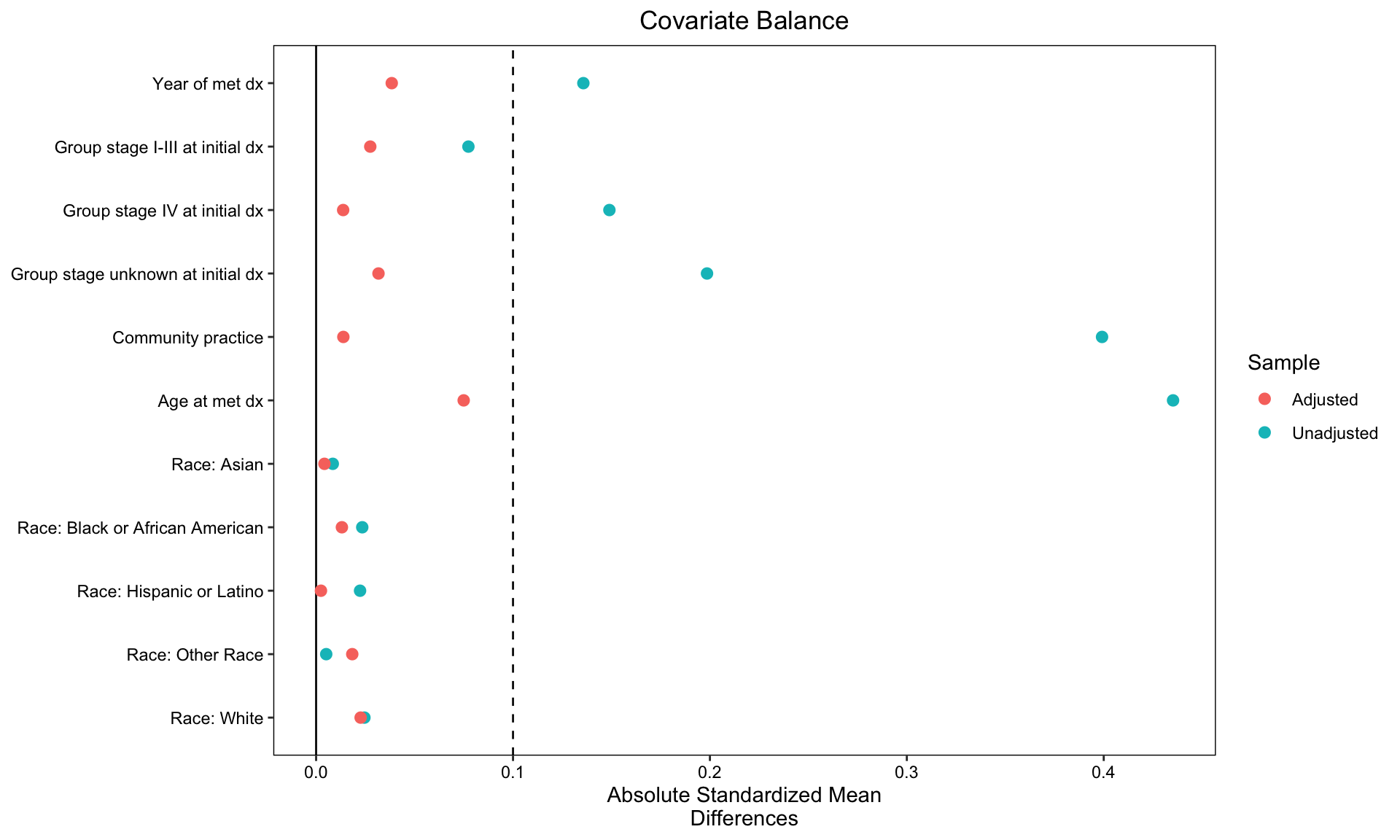}
    \caption{Covariate balance plot comparing FH and CGDB cohorts, displaying both weighted and unweighted absolute standardized mean differences. A threshold of 0.1 is used to indicate balance.}
    \label{fig:balance}
\end{figure}

We obtain a reference distribution of the required variables from the metastatic prostate cancer nationwide de-identified, EHR-derived Flatiron Health (FH) database. 
Unlike in the CGDB, an NGS test is not required for a patient to qualify for inclusion in this database. 
Therefore, the distribution of observed variables would not be truncated by test time. 
We apply the same inclusion-exclusion criteria to this database, and extract the adjustment variables from the resulting cohort. 
Then, we vertically concatenate the confounder datasets, labeling the FH database (non-truncated) observations as 1 and the CGDB (truncated) observations as 0. 
The density ratio weights are computed using the \texttt{WeightIt} R package, specifying inverse propensity weights targeting the ATT estimand, with the `treatment' being the truncated vs non-truncated cohort label. 
The weights for the CGDB patients are also multiplied by the sample size adjustment factor $\frac{\hat{P}(J=0)}{\hat{P}(J=1)}$; the weights for the FH patients are all 1.
The weighted distribution of confounders in the CGDB dataset is similar to the distribution in the FH dataset, as visualized in Figure \ref{fig:balance}.

Finally, we fit a weighted and risk set adjusted Kaplan-Meier estimator to the CGDB dataset, using the computed density ratio weights. 
This results in an estimated survival time of 16.5 months post-1L therapy start. Using the weighted nonparametric bootstrap with 1,000 resamples, we compute a 95\% confidence interval of [15.5, 17.8].
This suggests that the dependent left truncation led to a slightly upwardly biased estimate of survival in these patients, assuming that the non-truncated cohort would have the baseline confounder distribution as observed in the FH cohort. 
Under this assumption, we can interpret this estimate as the median survival time for the population of FH metastatic prostate cancer patients who would receive Foundation Medicine NGS testing.

\section{Discussion}

Real world EHR-derived datasets may be subject to left truncation, which can present complications for making statistical inferences on populations of scientific interest. 
Applying risk set adjustment in survival analyses is a straightforward solution, but requires independence between survival and cohort entry times. 
In the literature, testing for this condition is generally restricted to marginal dependence. 
If this is found to hold, then practitioners are faced with a lack of commonly accepted methods to analyze dependently truncated data.

In this work, we extend the conditions for valid inference beyond marginally independent left truncation by considering conditional independence. 
This occurs when confounding variables induce a spurious relationship between entry and survival times, and can be tested similarly as with marginal independence. 
Under conditionally independent left truncation, we show that certain parameters that are commonly of interest in real world evidence can be estimated unbiasedly. 
In particular, the methods we describe are complementary to analyses that compare real world cohorts to non-truncated trial cohorts, which we treat as reference data for learning the true confounder distribution. 
We demonstrate the effectiveness of our methods in simulation studies, and illustrate potential use by analyzing survival in a real world left truncated cohort. 
More broadly, these methods are applicable whenever the true distribution of confounders in the target population is known or estimable; as long as the truncated data can be appropriately weighted, the decomposition in Section 2.3 allows estimation of marginal parameters.

A limitation of this methodology is that it relies on the existence of baseline confounders yielding conditionally independent left truncation. 
In practice, these may be difficult to obtain. 
Using the CGDB as an example, it may be the case that NGS testing is usually performed for patients who have a worsening prognosis or cancer progression. 
These factors would not be measured at the start of a therapy, but would affect both survival and test timing. 
This type of treatment journey is particularly plausible for disease settings where NGS testing is not standard-of-care. 
An interesting future direction for this work would be to appropriately incorporate post-baseline or time-varying confounders. 
It would also be of interest to quantify how much bias remains when conditioning with a set of variables that reduces the dependence between survival and entry without fully eliminating it. 

In addition to the correct set of confounders, valid estimation and inference for marginal parameters also requires the weights to be well-estimated. 
To this end, standard diagnostics from the observational causal inference literature can be applied to assess the balancing quality of the weights. 
Moreover, when applying classifier-based density ratio estimation as we suggest in Section 2.3, cross-validation can also help to select an appropriate model class to train, since minimizing classifier loss directly corresponds to optimizing balance \citep{sondhi2020balanced}.

Despite these limitations, our work relaxes the necessary assumptions for valid inference under left truncation, allowing for a broader range of analyses to be conducted. 
We show that conditionally independent left truncation can be easily tested for, and results in unbiased estimates from common survival analyses. 
In practice, we recommend that researchers adopt this methodology when analyzing real world data subject to left truncation.

\clearpage

\bibliographystyle{apalike}
\bibliography{manuscriptrefs}

\newpage

\section{Appendix}

In this appendix, we sketch a proof of consistency for the weighted and risk set adjusted Kaplan-Meier estimator, given appropriate density ratio weights.
Recall that we define $T$ as the survival time, $C$ as the censoring time, and $E$ as the entry time. 
We observe $Y = \min(T, C)$ conditional on $Y > E$, with $\delta = I(T \le C)$ as the event indicator.
The observed event times are $x_j, j=1,...,m$.
$Z$ is a vector of confounders such that $Y \perp E | Z$

We begin by deriving the product form of the survival probability, given independently right-censored data.

\begin{align*}
    P(T \ge x_k) 
    &= \Pi_{j=0}^{k-1} S(x_{j+1}) \\
    &= \Pi_{j=0}^{k-1} P(T \ge x_{j+1} | T \ge x_j) \\
    &= \Pi_{j=0}^{k-1} [1 - P(T < x_{j+1} | T \ge x_j)] \\
    &= \Pi_{j=0}^{k-1} [1 - P(T \in [x_j, x_{j+1}) | T \ge x_j)] \\
    &= \Pi_{j=0}^{k-1} [1 - P(T = x_j | T \ge x_j)] \\
    &= \Pi_{j=0}^{k-1} [1 - P(Y = x_j, \delta = 1 | Y \ge x_j)] \\
    &= \Pi_{j=0}^{k-1} [1 - F(x_{j})]
\end{align*}

Then, the weighted and risk set adjusted Kaplan-Meier estimator for each term in this expression is given by
\begin{align*}
    \hat{F}(x_{j})
    &= \cfrac{\sum_{i=1}^n I(E_i \le x_j, Y_i = x_j) \delta_i w_i}{ \sum_{i=1}^n I(E_i \le x_j \le Y_i) w_i},
\end{align*}
where $w_i$ is a weight for subject $i$.
Specifically, 
$$
w_i = \cfrac{\pi(z_i)}{\pi(z_i|y_i > e_i)},
$$
the density ratio comparing the covariate distributions of the non-truncated and left truncated datasets.

To show consistency, we compute the expectations of the numerator and the denominator.
First the expectation of the numerator is:
\begin{align*}
    &\mathbbm{E}\left[ \sum_{i=1}^n I(E_i \le x_j, Y_i = x_j) \delta_i w_i | Y_i > E_i \right] \\
    &= \sum_{i=1}^n \mathbbm{E}[I(E_i \le x_j, Y_i = x_j) \delta_i w_i | Y_i > E_i] \\
    &= \sum_{i=1}^n \int I(E_i \le x_j, Y_i = x_j) \; \delta_i \; w_i \; \pi(y_i, e_i, z_i | y_i > e_i) \; d(y,e,z) \\
    &= \sum_{i=1}^n \int I(E_i \le x_j, Y_i = x_j) \; \delta_i \; w_i \; \pi(y_i, e_i | z_i, y_i > e_i) \; \pi(z_i | y_i > e_i) \; d(y,e,z) \\
    &= \sum_{i=1}^n \int I(E_i \le x_j, Y_i = x_j) \; \delta_i \; w_i \; \pi(y_i | z_i, y_i > e_i) \; \pi(e_i | z_i, y_i > e_i) \; \pi(z_i | y_i > e_i) \; d(y,e,z) \\
    &= \sum_{i=1}^n \int I(E_i \le x_j, Y_i = x_j) \; \delta_i \; w_i \; \pi(z_i | y_i > e_i) \; \pi(y_i | z_i) \; \pi(e_i | z_i) \; d(y,e,z) \\
    &= \sum_{i=1}^n \int I(E_i \le x_j, Y_i = x_j) \; \delta_i \; \pi(z_i) \; \pi(y_i | z_i) \; \pi(e_i | z_i) \; d(y,e,z) \\
    &= \sum_{i=1}^n \int I(E_i \le x_j, Y_i = x_j) \; \delta_i \; \pi(y_i, e_i, z_i) \; d(y,e,z) \\
    &= \sum_{i=1}^n \mathbbm{E}[I(E_i \le x_j, Y_i = x_j) \delta_i ] \\
    &= \sum_{i=1}^n \mathbbm{E}\left( \mathbbm{E}[I(E_i \le x_j, Y_i = x_j) \delta_i | Z] \right) \\
    &= \sum_{i=1}^n \mathbbm{E}\left( \mathbbm{E}[I(E_i \le x_j) I(Y_i = x_j) \delta_i | Z] \right) \\
    &= \sum_{i=1}^n \mathbbm{E}[I(E_i \le x_j) I(Y_i = x_j) \delta_i ] \\
    &= \sum_{i=1}^n P(E_i \le x_j)P(Y_i = x_j, \delta_i = 1) \\
    &= n P(E \le x_j)P(Y = x_j, \delta = 1)
\end{align*}

Similarly, for the denominator, we can obtain:
\begin{align*}
    &\mathbbm{E}\left[ \sum_{i=1}^n I(E_i \le x_j \le Y_i) \; w_i | Y_i > E_i \right] \\
    &= \sum_{i=1}^n \mathbbm{E}[I(E_i \le x_j \le Y_i) ] \\
    &= \sum_{i=1}^n \mathbbm{E}(\mathbbm{E}[I(E_i \le x_j \le Y_i) | Z ]) \\
    &= \sum_{i=1}^n \mathbbm{E}(\mathbbm{E}[I(E_i \le x_j) I(Y_i \ge x_j) | Z ]) \\
    &= \sum_{i=1}^n \mathbbm{E}[I(E_i \le x_j) I(Y_i \ge x_j)] \\
    &= n P(E_i \le x_j) P(Y_i \ge x_j)
\end{align*}

Therefore, by applying the continuous mapping theorem:
\begin{align*}
    \hat{F}(x_{j})
    &\longrightarrow \cfrac{n P(E \le x_j) P(Y = x_j, \delta = 1) }{n P(E \le x_j) P(Y \ge x_j)}
    = P(Y = x_j, \delta = 1 | Y \ge x_j),
\end{align*}
as required.

\end{document}